\documentclass[11pt,twoside]{article}
\usepackage{asp2006}
\usepackage{epsf}
\usepackage{psfig}
\usepackage{lscape}
\usepackage{graphicx}

\begin{document}

\title{Cosmological Parameters 2006}   

\author{Ariel G. S\'anchez\altaffilmark{1}, and Carlton M. Baugh\altaffilmark{2}}

\altaffiltext{1}{Instituto de Astronom\'{\i}a Te\'orica y Experimental (IATE), OAC, UNC, 
Laprida 854, X5000BGR, C\'ordoba, Argentina.
 }
\altaffiltext{2}{The Institute for Computational Cosmology, Department of Physics, University of Durham, South 
Road, Durham DH1 3LE, UK.}

\begin{abstract} 
Recent measurements of the cosmic microwave background radiation (CMB), particularly 
when combined with other datasets, have revolutionised our knowledge of the values of 
the basic cosmological parameters. Here we summarize the state of play at the end of 
2006, focusing on the combination of CMB measurements with the power spectrum of 
galaxy clustering.
We compare the constraints derived from the extant CMB 
data circa 2005 and the final 2dFGRS galaxy power spectrum, with the results obtained 
when the WMAP 1-year data is replaced by the 3-year measurements (hereafter WMAP1 and 
WMAP3). Remarkably, the picture has changed relatively little with the arrival of WMAP3, 
though some aspects have been brought into much sharper focus. One notable example of 
this is the index of primordial scalar fluctuations, $n_{\rm s}$. Prior to WMAP3, 
S\'anchez et~al. (2006) found that the scale invariant value of $n_{\rm s}=1$ was 
excluded at the 95\% level. With WMAP3, this becomes a $3\,\sigma$ result, with 
implications for models of inflation. We find some disagreement between the 
constraints on certain parameters when the 2dFGRS P(k) is replaced by the SDSS 
measurement. This suggests that more work is needed to understand the relation between 
the clustering of different types of galaxies and the linear perturbation theory 
prediction for the power spectrum of matter fluctuations.
\end{abstract}

\keywords{large scale structure of the universe - cosmic microwave background,cosmological parameters  }

\section{Introduction}
\label{sec:introduction}

In the last decade, the measurements of fluctuations in the temperature of the cosmic 
microwave background radiation (CMB) have shown a dramatic improvement marking the 
start of a new, data-rich era in cosmology (de Bernardis et al.~2000).

The CMB power spectrum encodes information about the values of the cosmological 
parameters. However, degeneracies exist between certain combinations of parameters 
which impose a limit on the precision attainable using CMB data alone.
In order to break these degeneracies it is necessary to combine the CMB data with other 
datasets, such as the power spectrum of galaxy clustering. The two-degree field galaxy 
redshift survey (2dFGRS) and the Sloan Digital Sky Survey (SDSS) are the largest 
redshift surveys available and give the most detailed description of the large scale 
structure of the Universe (LSS) as traced by galaxies. A consistency check between these two datasets 
would test the hypothesis about the relation between the galaxy power spectrum and the linear 
perturbation theory prediction for the power spectrum
of matter fluctuations.

The constraints quoted in this work were obtained by S\'anchez et~al.~(2006; hereafter 
S06). In Section~2, we describe the data used in the parameter 
estimation. In Section~3, we summarize the main results of S06 for the 
parameter constraints and perform a comparison with those obtained by Spergel 
et~al.~(2006).
We also compare the results obtained using the 2dFGRS and SDSS galaxy power spectra. 
Finally, we summarize our conclusions in Section~4.

\section{The Method}
\label{sec:method}

In order to constrain cosmological parameters, S06 used a compilation of recent 
measurements of the CMB, including the WMAP first year data (WMAP1; Hinshaw
et~al.~2003; Kogut et~al.~2003), extended to smaller angular scales with information from
 ACBAR (Kuo et~al.~2004), VSA (Dickinson et~al.~2004)and CBI (Readhead et~al.~2004).
These datasets were combined with the power spectrum of galaxy clustering 
measured from 
the final 2dFGRS catalogue by Cole et~al.~(2005). In order to model the effects of 
non-linearities and galaxy bias, S06 followed the scheme developed by Cole et~al. who 
applied a correction to the shape of $P(k)$ of the form
\begin{equation}
P_{\rm gal}(k)=b^{2}\frac{1+Qk^{2}}{1+Ak}\,P_{\rm lin}(k),  
\label{eq:q+a}
\end{equation}
where $A=1.4$ and $Q=4.6$ are the preferred values for the 2dFGRS and b is a constant 
bias factor. 
The power spectrum data was used 
for $0.02 \,h \mathrm{{Mpc^{-1}}}< k<0.15 \,h \mathrm{{Mpc^{-1}}}$.

S06 analysed a range of parameter sets and priors, allowing for massive neutrinos, 
curvature, tensors and general dark energy models. Here we focus on the `basic-six' 
parameter space, defined by
\begin{equation}
\mathbf{P^{b6} \equiv (}\omega_{\rm dm},\omega_{\rm b},\tau,n_{\rm s},A_{\rm s},\Theta), 
\label{eq:param6}
\end{equation}
from which other parameters can be derived.

\section{Results}
\label{sec:results}

\begin{table}[!t]
\caption{Comparison of the constraints on the basic six parameter set defined by (\ref{eq:param6}) from S\'anchez et~al.~(2006) and Spergel et~al.~(2006)}
\begin{center}{
\small
\begin{tabular}{cccc}
\tableline
\tableline
\noalign{\smallskip}
\noalign{\smallskip}
             & S\'anchez et al.~(2006) & \multicolumn{2}{c}{Spergel et al.~(2006)}\\
\noalign{\smallskip}
\tableline
\noalign{\smallskip}
\noalign{\smallskip} 
   Parameter & WMAP1(ext.)+2dFGRS      & WMAP3 only                &   WMAP3+2dFGRS \\
\noalign{\smallskip}
\tableline
\noalign{\smallskip}
$\omega_{\rm b}$  & $0.0225 \pm 0.0010$ &  $0.0223 \pm 0.0008 $  &  $0.0222 \pm 0.0007$ \\
$\omega_{\rm m}$  & $0.127 \pm 0.005$   & $0.126 \pm 0.009$      &  $0.1262 \pm 0.0048$ \\
$h$               & $0.735 \pm 0.022$   & $0.74 \pm 0.03$        &  $0.732 \pm 0.021$   \\
$\tau$            & $0.118 \pm 0.060$   & $0.093 \pm 0.029$      &  $0.083 \pm 0.029$   \\
$n_{\rm s}$       & $0.954 \pm 0.023$   & $0.961 \pm 0.017$      &  $0.948 \pm 0.016$   \\
$\sigma_{8}$      & $0.773 \pm 0.053$   & $0.76 \pm 0.05$        & $0.737 \pm 0.039$    \\
$\Omega_{\rm m}$  & $0.237 \pm 0.020$   & $0.234 \pm 0.035$      & $0.236 \pm 0.020$    \\
\noalign{\smallskip}
\tableline
\tableline
\end{tabular}
}\end{center}
\label{tab:params}
\end{table}

When the parameter space defined by (\ref{eq:param6}) is explored using CMB data alone, the results show a 
degeneracy that involves all six parameters and which is seen most clearly in $\tau$, $n_{\rm s}$ and 
$A_{\rm s}$. The 2dFGRS $P(k)$ helps to break this degeneracy, particularly by tightening the constraints on 
$w_{\rm dm}$: ($w_{\rm dm}=0.105^{+0.013}_{-0.013}$ for CMB data alone and 
$w_{\rm dm}=0.1051^{+0.0046}_{-0.0047}$ for CMB plus 2dFGRS).
When the CMB data is combined with the 2dFGRS $P(k)$, the recovered value of the spectral index of scalar 
perturbations is $n_{\rm s} = 0.954^{+0.023}_{-0.023}$ with $n_{\rm s} < 1$ at the $95\%$ confidence level, a 
deviation from scale invariance ($n_{\rm s}=1$). This result has strong implications for the inflationary 
paradigm.

Table \ref{tab:params} shows a comparison of the constrains on the basic-six parameter set obtained by S06 with 
those obtained by Spergel et al.~(2006) using information from the third year of flight of WMAP. The new WMAP 
data represent a significant improvement over WMAP1 due to the longer integration time and a better 
understanding of systematic effects. The new polarization data is particularly helpful in constraining $\tau$, 
breaking some of the degeneracies present in the temperature data. The two sets of constraints are in 
remarcable agreement. In particular, the new results confirm, with a higher statistical significance, the 
detection by S06 of a deviation from scale invariance.

\begin{figure}[t]
\centerline{\includegraphics[width=8cm]{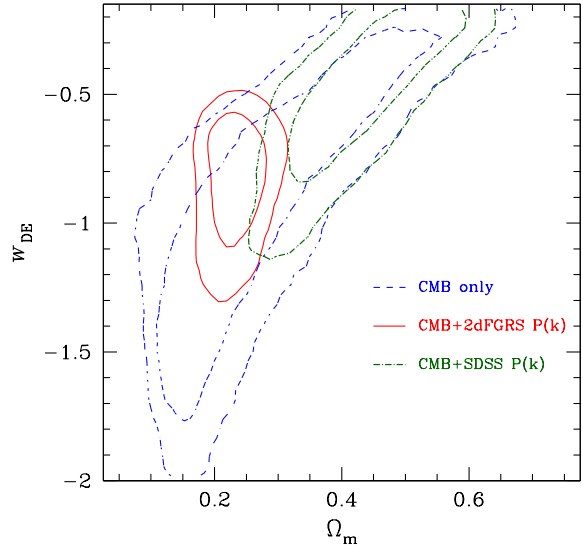}}
\caption{The marginalized posterior likelihood in the $\Omega_{\rm m}-w_{\rm DE}$ plane for the basic six plus 
$w_{\rm DE}$ parameter set. The lines show the 68\% and 95\% contours obtained in the CMB only (dashed), CMB+2dFGRS $P(k)$ (solid) and the CMB+SDSS $P(k)$ (dot-dashed) cases.}
\label{fig:b6_w}
\end{figure}

The marginalized constraints obtained for the CMB plus 2dFGRS $P(k)$ case are in striking agreement with those 
in the CMB only case, showing the impressive consistency between these datasets. There is a clear discrepancy, 
however, with the results obtained using CMB data plus the SDSS $P(k)$ estimated by Tegmark et~al.~(2004). 
This difference can be traced back to the density of dark matter, with the SDSS data pointing to 
higher values of $\omega_{\rm dm}$.

On considering the basic-six parameter space extended by the incorporation of the dark energy equation of 
state parameter, $w_{\rm DE}$, we find $w_{\rm DE}=-0.45_{-0.23}^{+0.23}$ using the SDSS $P(k)$, which is 
much higher than the value $w_{\rm DE}=-0.85_{-0.17}^{+0.18}$ obtained in the case of the 2dFGRS $P(k)$ and 
inconsistent with a cosmological constant. Again, this discrepancy is 
due to the preferred values of $\omega_{\rm dm}$. Fig.~\ref{fig:b6_w} shows the degeneracy in the 
$\Omega_{\rm m} - w_{\rm DE}$ plane for CMB data alone. Adding information from the galaxy power spectrum 
breaks this degeneracy. If the galaxy $P(k)$ data prefer a high value of $\Omega_{\rm m}$, as is the case for 
the SDSS data, then a high value of $w_{\rm DE}$ will result.

\section{Conclusions}
\label{sec:conclusions}

We have reviewed the constraints on the values of the basic cosmological parameters obtained using a 
combination of CMB data and the galaxy power spectrum of the final 2dFGRS. The data shows clear evidence for a 
departure from a scale invariant primordial spectrum of scalar fluctuations; the value $n_{\rm s}=1$ is 
formally excluded at the $95\%$ confidence level.

A comparison with the results from Spergel et al. (2006) obtained using the data from the third year of flight 
of the WMAP satellite shows an excellent agreement, which is a reassuring validation of the cosmological 
paradigm. In particular, the new results confirm the detection by S06 of a deviation from scale invariance 
with a higher statistical significance.

There is an impressive agreement between the results obtained for CMB data alone and for CMB data plus the 
2dFGRS power spectrum data. However, there is some tension between the constraints from the CMB and SDSS 
datasets. Cole, S\'anchez and Wilkins (2006) have shown that these differences are due to the $r$-band selected 
SDSS catalogue being dominated by more strongly clustered red galaxies, which
have a stronger scale dependent bias. It is therefore important to accurately model the distortion in the 
shape of the power spectrum caused by nonlinearity and scale dependent bias in order to obtain unbiased 
constraints on cosmological parameters from present and future galaxy surveys.

\end{document}